\documentclass[a4paper,fleqn,usenatbib]{mnras}
\usepackage{newtxtext,newtxmath}

\usepackage[T1]{fontenc}
\usepackage{ae,aecompl}

\usepackage[flushleft]{threeparttable}
\usepackage{media9}

\usepackage{graphicx}	
\usepackage{subcaption}
\usepackage{amsmath}	
\usepackage{amssymb}	

\title[Companions to Kepler detached eclipsing binaries]{Stability of planetary, single M dwarf, and binary star companions to Kepler detached eclipsing binaries and a possible five body system}

\author[Getley et al.]{
A. K. Getley,$^{1}$\thanks{E-mail: \href{mailto:alan.getley@usq.edu.au}{alan.getley@usq.edu.au}}
B. Carter,$^{1}$
R. King$^{1}$
and S. O'Toole$^{2}$
\\
$^{1}$University of Southern Queensland, Centre for Astrophysics, Toowoomba Qld 4350 Australia\\
$^{2}$Australian Astronomical Optics, Macquarie University, PO Box 915, North Ryde NSW 1670 Australia
}

\date{Accepted XXX. Received YYY; in original form ZZZ}

\pubyear{2020} 

\begin{document}
\label{firstpage}
\pagerange{\pageref{firstpage}--\pageref{lastpage}}
\maketitle

\begin{abstract}
In this study we identify 11 Kepler systems (KIC 5255552, 5653126, 5731312, 7670617, 7821010, 8023317, 10268809, 10296163, 11519226, 11558882 and 12356914) with a "flip-flop" effect in the eclipse timing variations O-C diagrams of the systems, report on what these systems have in common and whether these systems are dynamically stable. These systems have previously reported high eccentric binary stars with highly eccentric third bodies/outer companions. We find that all of the additional bodies in the system are dynamically stable for the configurations previously reported and are therefore likely to exist as described. We also provide additional evidence of KIC5255552 being a quadruple star system comprised of an eclipsing binary pair and non-eclipsing binary pair with the possibility of a fifth body in the system. With the advent of the NASA TESS exoplanet survey, its precision photometric monitoring offers an opportunity to help confirm more local eclipsing binary star companions, including planets.
\end{abstract}




\section{Introduction}

The `Kepler Eclipsing Binary Star Catalog' contains more than 2,000 eclipsing binary stars that have been observed during the Kepler mission \citep{prsa2011, slawson2011}. The high precision observations from Kepler enable eclipse time studies to be performed where variations in the eclipse times of binary stars can be used to detect third bodies \citep[e.g.][]{getley2017, borkovits2016}. Binary stars that have orbits aligned with the Earth will eclipse each other and detached and isolated binary stars should have eclipses that occur at predictable intervals. Plots of observed eclipse times (O) minus the calculated eclipse times (C), or O-C plots, may show variations from these predicted intervals. If these variations are also periodic it may be the result of a third body orbiting the binary stars \citep{beuermann2010}.

When performing an eclipse timing study on the eclipsing binaries contained in the Kepler catalogue, several O-C diagrams were found where the values begin to decrease, or increase, and then suddenly and rapidly reverse direction and change sign, i.e., eclipses that occur earlier than expected change to later than expected, or vice versa. The O-C curves for these systems then rapidly reverse sign again, or "flip-flop" (see Fig. ~\ref{fig:oc_diagram} for a visual example). The secondary eclipse O-C curve is out of phase with the primary eclipse O-C curve by a half orbital period. Examples of these "flip-flop" systems can be seen in \cite{borkovits2016}. Most of these systems also appear to have eclipse depth variations with differing magnitudes for each system. These systems all have similar reported eccentricities of the eclipsing binary as well as the highly eccentric orbit of the reported third body/outer companion. For the purposes of this paper, eclipsing binary is defined as the primary and secondary stars that eclipse each other i.e. producing the eclipses seen in the system O-C diagrams while third body/outer companion refers to one (or more) additional bodies orbiting the eclipsing binary.

The "flip-flop" features of the O-C diagrams and the high eccentricities raise the question of the dynamical stability of the systems and whether the systems with the reported configurations are stable. The dynamical stability of systems are important as outer companions in unstable orbits may result in the outer companion being ejected from the system within a short time period. However, stable orbits suggest the outer companion will remain within the system and are, therefore, more likely to exist as described and be observed \citep{horner2012a, horner2012b}. If an outer companion is stable for a range of configurations then the outer companion is more likely to exist as any detection errors won't have a dramatic effect on the determination of the stability of the system.

The aims of this study are to: perform a dynamical stability analysis on the systems found with highly eccentric binary star orbits and extremely high eccentric outer companion orbits; report on the source of the "flip-flop" effect and the stability of the systems KIC5255552, KIC5653126, KIC5731312, KIC7670617, KIC7821010, KIC8023317, KIC10268809, KIC10296163, KIC11519226, KIC11558882 and KIC12356914 with the proposed third bodies; comment on the likelihood of these proposed third bodies existing; and comment on the likelihood of more of these "flip-flop" systems existing that continue to go undetected.

\section{Method}
\label{section: method}
The Kepler data were used to produce O - C diagrams for detached eclipsing binaries to study eclipse timing variations. We created a C++ program, called {\sc bet} or {\sc binary eclipse timings}, to determine the mid-eclipse times of as many primary and secondary eclipses in the Kepler detached binary systems as possible \citep[see][]{getley2017}. {\sc bet} is based on the software {\sc transit analysis package} \citep{gazak2012} which uses the analytic formulae from \cite{mandel2002}. The analytic formulae describe a system of two objects, using parameters including orbital period, radius ratio of the two objects, mid-eclipse time, orbital inclination and eccentricity, during various points throughout an orbit. The O-C diagrams of the Kepler systems shown in this paper were created using {\sc bet} and found to contain rapid variations with the primary and secondary eclipse O-C curves out of phase.

{\sc rebound} is an N-body integrator with Python and C implementations \citep{rein2015}. Systems of bodies are able to be set up and integrated over time to estimate the orbital characteristics, such as semi-major axis and eccentricity, at various intervals. By simulating the positions and the evolution of the estimated orbital characteristics of a system over a long time period we can determine if the proposed system is in a stable orbit (allowing it to have been observed) or if it is in an unstable orbit and likely to eject one or more of the bodies. Eclipse times were obtained from the simulation and an O-C diagram produced to make sure that the distinctive characteristics of the actual O-C diagrams were present. The systems with these orbital characteristics were also integrated for $10^6$ years. These same systems were then integrated again 40 times for $10^4$ years with random values for the mean longitude, argument of pericentre and longitude of ascending node of the orbit of the outer companion and eclipsing binary. The purpose of the random values was to see if the third bodies were stable in this very specific configuration or if third bodies were stable for a range of configurations.

For the {\sc rebound} models used, the value for the longitude of the ascending node for the eclipsing binary (i.e. $\Omega_{binary}$) was fixed to 90\textdegree. We found when it was fixed to 0\textdegree\ that although the "flip-flop" features of the O-C diagrams still occurred, the primary and secondary eclipse O-C curves were in phase rather than out of phase as seen within the real O-C diagrams. The value for the longitude of the ascending node for the outer companion (i.e. $\Omega_{companion}$) was set such that $\Omega_{companion} = \Omega_{binary} + \Delta\Omega$. 

The individual masses for the primary and secondary star were calculated using the sum of the masses in \cite{borkovits2016} and the temperature of the systems. Making the assumption that the primary star significantly dominates the temperature of the system we can search for the corresponding mass of a star at that temperature from \cite{pecaut2013}\footnote{ With additional details from \url{http://www.pas.rochester.edu/~emamajek/EEM_dwarf_UBVIJHK_colors_Teff.txt} }. This becomes the estimate for the mass of the primary star. Using either a calculated mass ratio or a sum of masses of the primary and secondary star with the estimated primary star mass, calculating the estimated mass of the secondary star becomes trivial. Finally, we compare the J-H colour/magnitude difference of the system with the estimate for the primary star in order to perform a check on the assumption that the primary star significantly dominates the system. We tested this process for estimating masses against Kepler systems with known masses for the primary and secondary stars, Kepler-16 \citep{doyle2011}, Kepler-34 and 35 \citep{welsh2012}, Kepler-38 \citep{orosz2012}, Kepler-47 \citep{orosz2012_2}, all with at least one confirmed planet. Our estimates  for the primary and secondary masses agree with the reported masses within $\sim$10\% or less. We also tested against systems with no confirmed outer companions, KIC9851142 \citep{cakirli2015} and KIC1571511 \citep{ofir2012}, and found our mass estimates agreed with the reported masses within $\sim$20\% or less. 

\begin{table*}
  \centering
  \caption{A list of orbital properties for the systems used in the dynamical stability studies. Values from \protect\cite{borkovits2016}. With orbital periods for the third bodies less than, or approximately equal to, the Kepler viewing window the orbital periods of the third bodies and their properties will be likely to reflect the true nature of the systems.}
  \label{tab:systems_shortperiod}
    \begin{tabular}{lccccccccccc}
      \hline
      KIC No. & $P_1$ & $P_2$ & $m_{a+b}$ & $m_c$ & $e_1$ & $e_2$ & $i_1$ & $i_2$ & $\omega_1$ & $\omega_2$ & $\Delta\Omega$\\
       & (Days) & (Days) & ($M_\odot$) &  & & & (deg) & (deg) & (deg) & (deg) & (deg)\\
      \hline
      5255552 & 32.465339 & 862.1 & 1.7 & 0.7$M_\odot$ & 0.30668 & 0.4342 & 83.8 & 89.5 & 105.27 & 37.3 & -2.8\\
      5653126 & 38.49233 & 968 & 1.8 & 1.1$M_\odot$ & 0.247 & 0.189 & 87 & 78 & 313 & 326 & -5 \\
      5731312 & 7.9464246 & 911 & 1.1  & 0.13$M_\odot$ & 0.4196 & 0.584 & 88.5 & 77.3 & 183.9 & 25.9 & 36.4 \\
      7821010 & 24.2382191 & 991 & 2.3 & 2.6$M_{Jup}$ & 0.6791 & 0.372 & 88 & 105 & 239.234 & 126 & -19 \\
      8023317 & 16.57907 & 610.6 & 1.3 & 0.15$M_\odot$ & 0.2511 & 0.249 & 88 & 93 & 177.7 & 164 & -49.3 \\
      11519226 & 22.161767 & 1437 & 1.44 & 1.25$M_\odot$ & 0.18718 & 0.332 & 88 & 89 & 358.4 & 321.7 & 17.0 \\
      \hline
    \end{tabular}
\end{table*}

\begin{table*}
  \centering
  \caption{A list of orbital properties for the systems used in the dynamical stability studies. With orbital periods for the third bodies larger than the Kepler viewing window the ability to accurately resolve these properties is difficult, however they still give an indication of the possible configuration of these systems. Values from \protect\cite{borkovits2016}.}
  \label{tab:systems_longperiod}
  \begin{tabular}{lccccccccccc}
      \hline
      KIC No. & $P_1$ & $P_2$ & $m_{a+b}$ & $m_c$ & $e_1$ & $e_2$ & $i_1$ & $i_2$ & $\omega_1$ & $\omega_2$ & $\Delta\Omega$ \\
       & (Days) & (Days) & ($M_\odot$) &  & & & (deg) & (deg) & (deg) & (deg) & (deg) \\
      \hline
      7670617 & 27.70317 & 3304 & 0.9 & 0.55$M_\odot$ & 0.249 & 0.707 & 86 & 89 & 135 & 86.4 & -147.8 \\
      10268809 & 24.70843 & 7000 & 1.5 & 1.4$M_\odot$ & 0.314 & 0.737 & 84 & 94 & 143.1 & 292.6 & 21.6 \\
      10296163 & 9.296847 & 15271 & 1.4 & 0.5$M_\odot$ & 0.354 & 0.73 & 86 & 127 & 45.7 & 355 & -40 \\
      11558882 & 73.9135 & 4050 & 1.9 & 0.4$M_\odot$ & 0.365 & 0.30 & 88 & 84 & 169 & 105 & -43 \\
      12356914 & 27.3083183 & 1804 & 1.8 & 0.41$M_\odot$ & 0.325 & 0.385 & 88 & 60 & 113.2 & 36.5 & -30.4 \\
      \hline
    \end{tabular}
\end{table*}

\begin{figure}
    \includegraphics[width=\columnwidth]{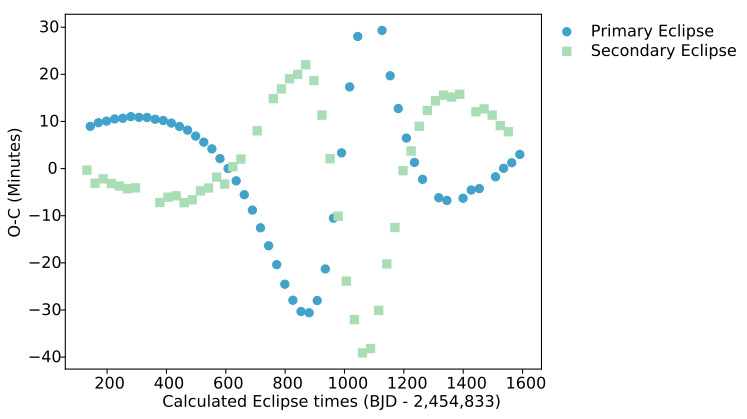}
    \caption{Observed minus Calculated (O-C) diagram of KIC12356914 showing the sudden and rapid period flip in the primary (blue circles) and secondary (green squares) eclipses. For example, at $\sim$900 days the primary eclipses go from occurring $\sim$30 minutes earlier than calculated to $\sim$30 minutes later than calculated in the space of $\sim$200 days.}
    \label{fig:oc_diagram}
\end{figure}

The first systems to be selected for the dynamical stability study were KIC5255552, KIC5731312, KIC7670617, KIC10268809, KIC12356914 as these systems were identified as part of our own eclipse time study of the Kepler eclipsing binaries that had matching entries in \cite{borkovits2016}. These systems all contained a unique "flip-flop" feature or sudden period change in their O-C diagrams as seen in Fig. ~\ref{fig:oc_diagram}. The inferred properties of these systems were compared to see what all the systems had in common. The systems were found to have binary eccentricities ranging between $\sim0.25$ and $\sim0.42$ and third bodies with eccentricities of at least 0.385. The rest of the systems in \cite{borkovits2016} were checked to see if there were any other systems that matched these criteria. Finally, the O-C diagrams of the systems were visually compared to find other possible candidates. The complete list of systems and their orbital properties can be found in Table ~\ref{tab:systems_shortperiod} and Table ~\ref{tab:systems_longperiod}. Two systems, KIC4055092 and KIC9715925, were found to match the selection criteria, however these systems were not a part of the dynamical stability study as both of these systems have mass estimates for the primary star that exceed the mass estimates for the total system. Another two systems, KIC6794131 and KIC7177553, were also possible candidates for the dynamical stability study, however accurate values for $m_{a+b}$ were not obtainable from \cite{borkovits2016}. As such, reliable models in {\sc rebound} were unable to be made for these four systems and they were not included in the study.

The systems in Table ~\ref{tab:systems_longperiod} are listed separately due to the long period nature of the outer companions. These third bodies all have periods longer than the window of Kepler's observations, and so, while models and fits can give us an indication of the properties and type of third bodies located within the systems, the margin of error in the values are likely too great to make firm conclusions. We can expect any estimate of the orbital period to be a lower limit due to the uncertainty involved in observing a system for less than one complete orbital period. The outer companion mass is likely to be an upper limit as lower masses are more detectable at longer periods \citep{watson2010}. For those systems with orbital periods less than the period of Kepler's observations the values listed in Table ~\ref{tab:systems_shortperiod} are likely to be accurate with a smaller margin of error.

\begin{table}
  \centering
  \caption{Additional information about the systems found in Table ~\protect\ref{tab:systems_shortperiod} and Table ~\protect\ref{tab:systems_longperiod}. The temperature of the system comes from the Kepler Eclipsing Binary catalogue. ~\protect\cite{pecaut2013} and the temperature is used to estimate the primary star mass and, with the values of $m_{a+b}$ from Table ~\protect\ref{tab:systems_shortperiod} and Table ~\protect\ref{tab:systems_longperiod}, the secondary star mass.}
  \begin{threeparttable}
    \begin{tabular}{lcccc}
      KIC No. & Temperature & $m_a$ & $m_b$\\
       & K & ($M_\odot$) & ($M_\odot$)\\
      \hline
      5255552 & 4775 & 0.96\tnote{1} & 0.74\\
      5653126 & 5766 & 1.02 & 0.78 \\
      5731312 & 4658 & 0.73 & 0.37\\
      7821010 & 6298 & 1.23 & 1.07\\
      8023317 & 5625 & 0.98 & 0.32\\
      11519226 & 5646 & 0.98 & 0.46 \\
      7670617 & 4876 & 0.75 & 0.15\\
      10268809 & 5787 & 1.07 & 0.43 \\
      10296163 & 6229 & 1.21 & 0.19 \\
      11558882 & 6066 & 1.14 & 0.76 \\
      12356914 & 5368 & 0.90 & 0.90 \\
      \hline
    \end{tabular}
    \begin{tablenotes}
      \item [1] Mass of the primary star is larger than would be expected from the temperature of the system, though the total mass of the binary star system matches and is expected to be useful to determine the stability of the outer companion.
    \end{tablenotes}
  \end{threeparttable}
  \label{tab:masses}
\end{table}

The primary and secondary masses for the systems listed in Table \ref{tab:systems_shortperiod} and Table \ref{tab:systems_longperiod} were calculated as described and are listed in Table \ref{tab:masses}. A number of the systems in Table ~\ref{tab:systems_shortperiod} and Table ~\ref{tab:systems_longperiod} have outer companion masses that are almost as large, or even larger, than one or both of the stars in the binary system. If these third bodies significantly contribute to the flux of the system then the individual mass for the primary star would be larger than estimated and the mass for the secondary star would be lower (although the total mass of the binary system would be unaffected).

With the systems set up in {\sc rebound} and integrated, plots are produced showing eccentricity vs time and semi-major axis vs time. By considering these plots we are able to view the evolution of the system over the defined period and determine whether any object is likely to be ejected from the system. For example, by considering the change in semi-major axis we can tell if an outer companion stays within the system or is moving further away from the binary stars and being ejected out of the system.

The light curves for the systems with inclinations of close to 90\textdegree\ were also visually inspected to look for any additional eclipsing events. Additional eclipsing events are a direct way of confirming the existence of additional bodies and may provide additional information about the characteristics and orbital properties of any additional bodies.

\section{Results}

\begin{figure}
    \includegraphics[width=\columnwidth]{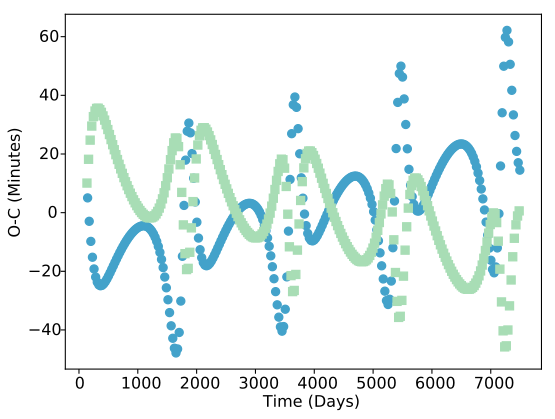}
    \caption{Simulated Observed minus Calculated (O-C) diagram of KIC12356914 showing the sudden and rapid period flip in the primary (blue circles) and secondary (green squares) eclipses like the sudden "flip-flops" seen in Fig. ~\protect\ref{fig:oc_diagram}.}
    \label{fig:sim_oc_diagram}
\end{figure}

The Kepler "flip-flop" systems appear visually unique upon the first consideration of their O-C diagram (Fig. ~\ref{fig:oc_diagram}). The primary and secondary eclipses O-C variations are out of phase with each other, and there are sharp and rapid "flip-flops" indicating eclipses rapidly transitioning from earlier than expected to later than expected (or vice versa). An example of a simulated model's O-C diagram can be seen in Fig. ~\ref{fig:sim_oc_diagram}. The simulated O-C diagram shows the same out of phase and rapid variations that can be seen in the actual O-C diagrams from observed data.

\begin{figure}
    \includegraphics[width=\columnwidth]{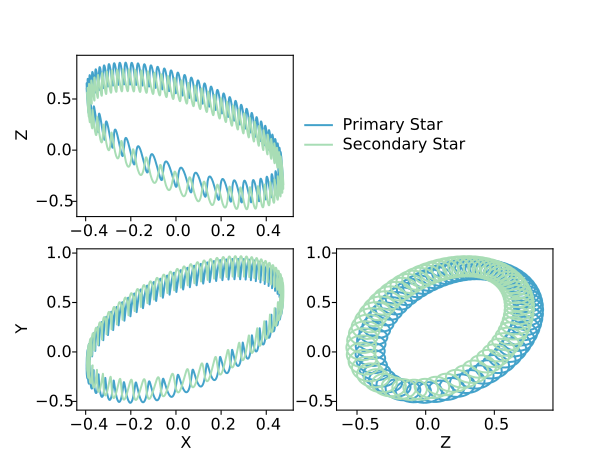}
    \caption{Plot of the XYZ coordinates of the two stars in the eclipsing binary of KIC12356914 showing a wobble around the centre of mass of the systems and the apsidal precession (particularly noticeable in the YZ plot) throughout a single orbit of the outer companion. Note: Animations of the binary star and outer body orbits will be available online as supplementary material. The observer is in the positive X direction with the Y axis running horizontal and the Z axis vertical.}
    \label{fig:binaryslices}
\end{figure}

\begin{figure}
    \includegraphics[width=\columnwidth]{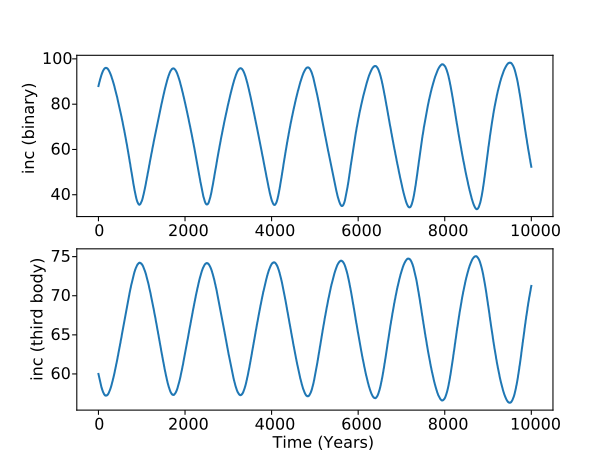}
    \caption{The change in inclination of the eclipsing binary (top) and third body (bottom) for the system KIC12356914 over $10^4$ years. The eclipsing binary inclination changes between $\sim$40\textdegree\ and $\sim$100\textdegree. As a result, there are likely to be extended intervals of time when no eclipses of the eclipsing binary will be seen from Earth.}
    \label{fig:inclinationevolution}
\end{figure}

The models from {\sc rebound} allowed us to produce visual representations of the bodies and their orbits within the systems found in Table ~\ref{tab:systems_shortperiod} and Table ~\ref{tab:systems_longperiod}. By producing visual representations of the binary star orbits (Fig. ~\ref{fig:binaryslices}), animating the binary star and outer companion orbits and the inclination evolution of the systems (Fig. ~\ref{fig:inclinationevolution}) we were able to determine that all systems with the "flip-flop" O-C variations exhibit similar behaviour/orbital configurations as described in Section \ref{section: method}. The binary stars are locally bound together and both orbit and exhibit apsidal precession around the centre of mass of the entire system. The period of the eclipsing binary apsidal precession around the centre of mass appears to be the same as the orbital period of the outer companion, likely due to the dynamical interactions between the outer body, primary and secondary stars. The third bodies orbit the centre of mass opposite the binary stars. The orbits of the binary stars and the system as a whole are provided as animations available as additional supplementary material online. The models provide clarity on the orbits of the bodies within the system and explain the features seen in the O-C diagrams.

\begin{figure*}
  \begin{subfigure}{\columnwidth}
    \includegraphics[width=\columnwidth]{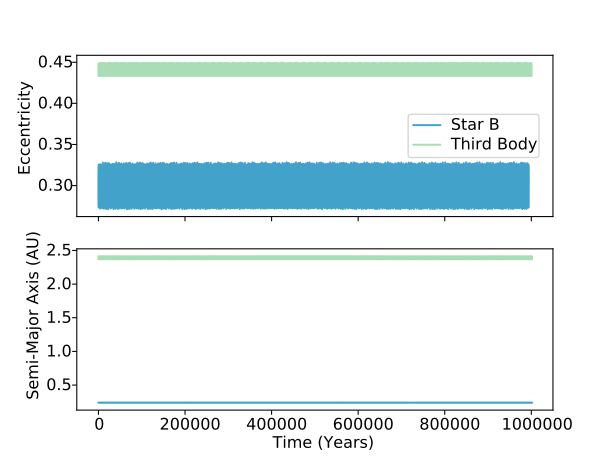}
    \caption{KIC5255552}
    \label{fig:5255552_long}
  \end{subfigure}
  \begin{subfigure}{\columnwidth}
    \includegraphics[width=\columnwidth]{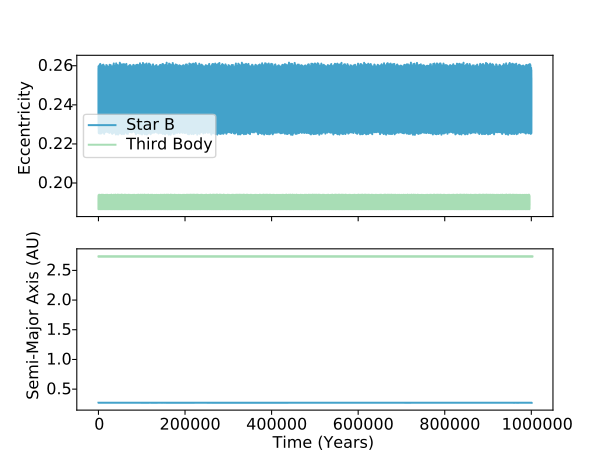}
    \caption{KIC5653126}
    \label{fig:5653126_long}
  \end{subfigure}
  \begin{subfigure}{\columnwidth}
    \includegraphics[width=\columnwidth]{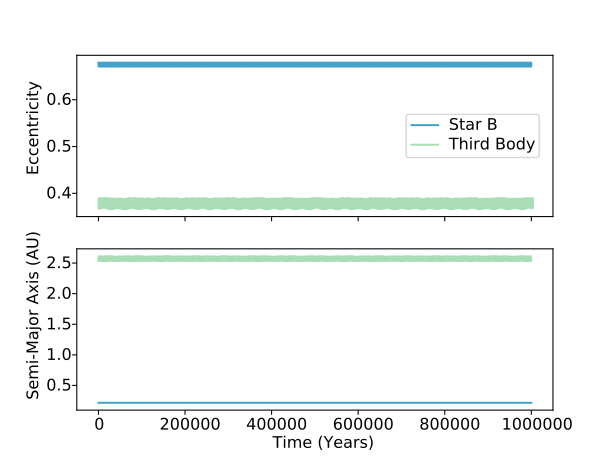}
    \caption{KIC7821010}
    \label{fig:7821010_long}
  \end{subfigure}
  \begin{subfigure}{\columnwidth}
    \includegraphics[width=\columnwidth]{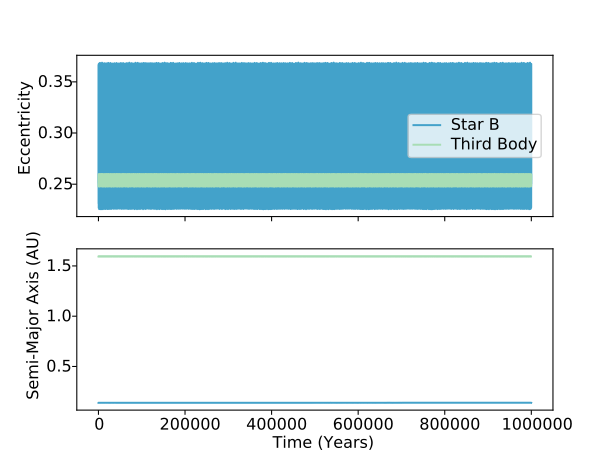}
    \caption{KIC8023317}
    \label{fig:8023317_long}
  \end{subfigure}
  \begin{subfigure}{\columnwidth}
    \includegraphics[width=\columnwidth]{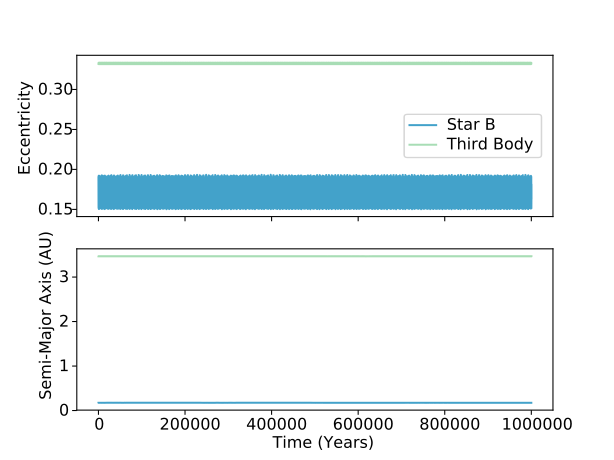}
    \caption{KIC11519226}
    \label{fig:11519226_long}
  \end{subfigure}
  \begin{subfigure}{\columnwidth}
    \includegraphics[width=\columnwidth]{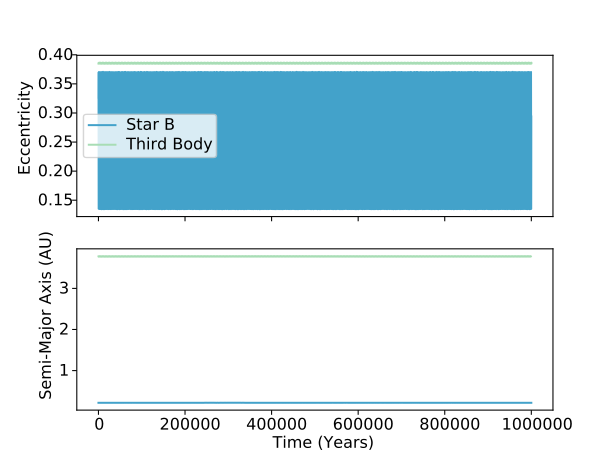}
    \caption{KIC12356914}
    \label{fig:12356914_long}
  \end{subfigure}
  \caption{Eccentricity and semi-major axis of the secondary star and third body/outer companion after integration in {\sc rebound} for a period of $10^6$ years for the systems listed in Table ~\ref{tab:systems_shortperiod} and Table ~\ref{tab:systems_longperiod}. Note: Figures for additional systems will be available online as supplementary material.}
  \label{fig:allfig}
\end{figure*}

The results of integrating the systems for $10^6$ years can be seen in Figures ~\ref{fig:5255552_long} to ~\ref{fig:12356914_long}. All of these systems were found to be stable over $10^6$ years. The eccentricities of the eclipsing binary combined with the high eccentricities of the outer companion do not appear to compromise the long-term stability of the systems. While the eccentricities of the objects in the systems varied over differing time-scales and by differing amounts, the semi-major axis remained relatively constant and, therefore, the outer companions remained within each system. As illustrated by Fig. ~\ref{fig:8023317_long} and Fig. ~\ref{fig:12356914_long}, while the eccentricity of the binary stars can vary significantly, this did not necessarily translate to a major change in eccentricity of the outer companion or the semi-major axis of the system. The systems were also found to be stable for $10^4$ years when random values were used for the mean longitude, argument of pericentre and longitude of pericentre of the outer companion and the eclipsing binary. This increases the likelihood of the outer companions existing as slight changes or deviations from the proposed orbital properties still produced stable orbits.

\section{Discussion}

\begin{figure}
    \includegraphics[width=\columnwidth]{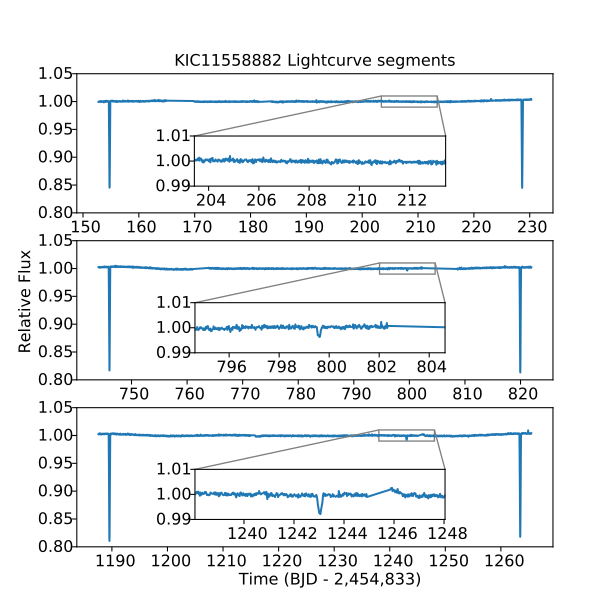}
    \caption{Secondary eclipses for KIC11558882 are not initially viewable. However, as time progresses and the inclination/binary star orientation changes secondary eclipses come in to view.}
    \label{fig:appearingeclipses}
\end{figure}

The out of phase variations in the O-C diagrams for primary and secondary eclipses are likely the result of apsidal motion \citep{zasche2015}. The apsidal motion and rapid eclipse time transitions are features that appear in all of the O-C diagrams of the models when an outer companion as described in Table ~\ref{tab:systems_shortperiod} or Table ~\ref{tab:systems_longperiod} is present. The light curves of some of these systems also show significant eclipse depth variations. The eclipse depth variations are likely due to the dynamics of the system at play due to apsidal and nodal precession \citep{kane2012}, and the evolution of the inclination in the system over time. Apsidal motion and nodal precession are illustrated in the simulated orbits in Fig. ~\ref{fig:binaryslices} while inclination evolution over time for a system can be seen in Fig. ~\ref{fig:inclinationevolution}. Inclination evolution does not necessarily only change the depth of the eclipses seen but also whether we see the eclipses at all. For example, secondary eclipses for KIC11558882 are not initially seen in the light curve but begin to appear around 800 days (BJD - 2,454,833) and remain for the rest of the observing window (Fig. ~\ref{fig:appearingeclipses}).

The Kepler mission viewed these systems for approximately 1400 days \citep{conroy2014}, and it is fortunate that the observation period of Kepler coincided with the point in the outer companion's orbit that results in the sudden "flip-flop" nature of the period changes. For third bodies that have orbital periods greater than 1400 days, part of the orbit will be unobserved and the "flip-flop" effect potentially missed. The greater the orbital period of the outer companion, the greater the chance of missing this dynamical effect in the observations. The sudden period changes are so rapid, some occurring over approximately 100 days, that even an orbital period of $\sim$1700 days could result in this system characteristic going undetected in the Kepler data.

The set of orbital properties within a system jointly influence the potential for transits or eclipses to be seen in the light curve. The probability of a transit occurring decreases as the orbital period increases ~\citep{kane2008} so while KIC10268809, for example, has inclinations that may indicate the possibility of transits (84\textdegree\ and 94\textdegree\ for the binary stars and outer companion respectively), the very long orbital period of the outer companion results in transits being unlikely to occur. Extra events can be seen in the light curve of KIC5255552, indicating that transits occur, and there are also additional eclipses that a third body may not account for, thus indicating the possibility of a quadruple system \citep{zhang2018}. None of the other systems considered in this study have definite or clear additional events occurring within the light curve, however it is possible KIC11519226 contains an additional eclipse (described in section \ref{kic11519226}). The equation for the probability of a third body transit/eclipse being seen from Earth is 

\begin{equation}
	P_{tr} = 0.0045 ({1AU\over a})({R_\star + R_{} \over R_\odot})[{1 + e \cos({\pi \over 2} - \omega) \over 1 - e^2}]
	\label{eq:transitprobability}
\end{equation} 

where a is the semi-major axis, e is the eccentricity and $\omega$ is the longitude of periastron of the third body and and the orientation of the orbit of the third body is assumed to be random \citep{charbonneau2006}. Using equation \ref{eq:transitprobability} and the mean radius of stars from \cite{pecaut2013} with the masses and other orbital characteristics in Table ~\ref{tab:systems_shortperiod} we can calculate the probabilities of seeing transits from the systems with third bodies. We find that the probability of extra events occurring in KIC5255552, KIC8023317 and KIC11519226 to be less than 1\% and that the extra events seen in KIC5255552 must be due to an extremely fortuitous occurrence.

In some systems, the sudden period flip in the O-C diagram may be the only indication of the presence of an outer companion. It is likely, given the large number of eclipsing binary stars observed with Kepler, that there are a number of systems that have been observed and classified as not containing an outer companion when in actuality the observations of Kepler haven't been long enough to observe the effects of an outer companion. With only 11 systems displaying the "flip-flop" behaviour out of the more than 2000 Kepler eclipsing binary systems and almost half of the systems having an outer companion reported with greater than a $\sim$1400 day orbital period, it is likely that there are many more systems that have outer companions that remain undetected due to orbital configurations that did not result in notable O-C diagrams within the Kepler viewing window. The approximately 1400 day viewing window of Kepler will necessarily bias the detection results to systems that have outer companions with orbital periods of less than 1400 days. As Tables ~\ref{tab:systems_shortperiod} and ~\ref{tab:systems_longperiod} contain a similar number of systems it is possible, if not likely, that the "flip-flop" characteristic seen in the O-C diagrams will exist in a wide range of systems that have already been observed but not during this "flip-flop" window.

All of the systems in Table ~\ref{tab:systems_shortperiod} and Table ~\ref{tab:systems_longperiod} were integrated 40 times each for $10^4$ years with random initial values for the mean longitude, argument of pericentre and longitude of pericentre of the third body orbit and the eclipsing binary. While the random values can produce systems with O-C diagrams that vary significantly from the previously calculated values, the systems are still found to be stable. This exercise shows that even for a wide range of (though not necessarily all) orbital configurations systems with these mass and eccentricity values are likely to be stable.

\subsection{KIC5255552}

The KIC5255552 reported outer companion mass of 0.7$M_\odot$ \citep{borkovits2016} closely matches the estimated mass of the secondary star at 0.74$M_\odot$. If this system were to contain a similar tertiary star to the secondary star we would expect this to have an effect on the system, for example, in the reported colours of the system and therefore affect mass estimates. KIC5255552 has a 2MASS J-H magnitude difference of 0.507 which approximately matches a K3V star \citep{pecaut2013}. Larger mass dwarf stars will have a smaller J-H magnitude difference, while smaller mass stars have a larger J-H value. If the mass of the outer companion was as large or larger than that of the secondary star, we would expect a smaller J-H magnitude difference, and therefore earlier spectral type. The estimated primary star mass was higher than expected from the temperature of the system and it is possible the J-H magnitude difference indicating a K3V star with a mass of 0.75$M_\odot$ more accurately reflects the primary star mass.

\begin{figure}
    \includegraphics[width=\columnwidth]{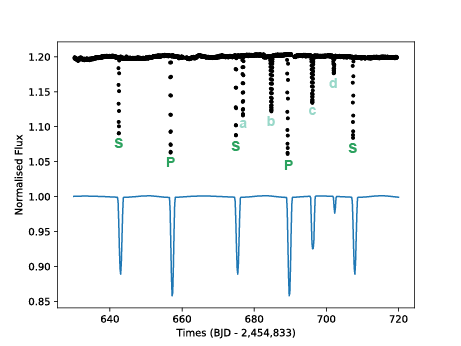}
    \caption{Top (Black): A vertically shifted segment of the KIC5255552 light curve showing the regular primary (P) and secondary (S) eclipses as well as the additional eclipsing events (a, b, c and d). Bottom (Blue): A segment of the modelled light curve of KIC5255552 eclipsing binary system with a single companion as described. While there are extra eclipsing events (corresponding to events c and d in the actual light curve), a single companion does not account for the a and b eclipsing events.}
    \label{fig:5255552_lc}
\end{figure}

KIC5255552 is unique amongst all of the systems considered in this study as it showed clear eclipsing events that cannot be attributed to the binary star alone. The light curve of KIC5255552 has a number of groups of extra eclipsing events, one group is shown in the top plot of Fig. \ref{fig:5255552_lc}. Four extra observed eclipses (a, b, c and d) can be seen in this group. This system was then modelled using {\sc phoebe} or {\sc PHysics Of Eclipsing BinariEs} \citep{horvat2018} with the binary stars and a third body as described in Table ~\ref{tab:systems_shortperiod}. However, only two additional eclipsing events can be seen in the modelled light curve in the bottom section of Fig. ~\ref{fig:5255552_lc}, corresponding to eclipses c and d seen in the actual light curve. The number of observed eclipsing events indicates that KIC5255552 contains a fourth body while the grouping of eclipsing events suggests that the third and fourth body are themselves in a binary star configuration. As no eclipses from the companion binary star are seen in the light curve we interpret this system as a non-eclipsing binary which itself eclipses an eclipsing binary.

There are clear additional groups of eclipsing events located around approximately 690 and 1542 days, representing the eclipsing binary passing in front of the companion binary and 948 days, representing the eclipsing binary passing behind the companion binary;  \citep{zhang2018}. A particularly large eclipsing event occurs at approximately 1548 days and is expected to be the primary star of the eclipsing binary blocking the light from both stars of the companion binary. \cite{zhang2018} note a possible additional eclipsing event occurs at approximately 1278 days; however it is a very shallow and isolated event. It is possible other events occurred slightly earlier than this event. However, they correspond to a time when no observations were taken. Given the probable binary nature of the companion if this is an independent, physical, eclipsing event it may indicate the presence of a fifth body in the system rather than a fourth body suggested by \cite{zhang2018}. 

\subsection{KIC5653126} \label{kic5653126}

The mass of the outer companion around KIC5653126 is reported to be 1.1$M_\odot$ \citep{borkovits2016}. Using the method described in section \ref{section: method}, we estimate the masses of the eclipsing binary primary and secondary stars to be 1.02$M_\odot$ and 0.78$M_\odot$ respectively. 

The 2MASS J-H magnitude difference of KIC5653126 is 0.247 which approximately matches an F9.5V star which is consistent with the mass of the reported outer companion. This may be because the outer companion is a single star that dominates the J-H colour of the system. The presence of significant third light can result in unreliable mass ratio determinations \citep{hambalek2013}. As a result the mass estimates for the primary and secondary stars of the eclipsing binary would not be accurately determined. Alternatively, the primary star of the eclipsing binary may dominate the temperature of the system with the outer companion contributing only slightly to the J-H colour of the system. However, assuming relatively accurate combined mass estimates, in either case the outcome of the stability check performed would remain the same.

In the second case, if the outer companion contributes slightly to the J-H colour of the system, it is possible that the outer companion is itself an additional binary rather than a single star companion. As there are no additional eclipsing events seen in the light curve of KIC5653126 this potential additional binary is unlikely to be eclipsing, nor is it likely that a star in either the eclipsing binary or this potential companion binary eclipses a star in the other binary. This is the expected result with the inclination of the outer companion being 78\textdegree.

\subsection{KIC7821010}

Another system of note is KIC7821010 which has a third body mass of just $\sim$2.6 Jupiter masses \citep{borkovits2016}. The evidence for this third body mass (i.e. the eclipse timing fit, the models reproducing the O-C effects and the stability of the system) all strongly point to the existence and viability of this as a planetary candidate. The third body in this system is in an orbit with an inclination of 105\textdegree\ and with a configuration similar to that of the planetary mass third body found orbiting KIC5095269 \citep{getley2017}. It is also further evidence that low mass objects can have a significant effect on the orbital properties of the host stars and also that, for at least some orbital configurations, eclipse timing variations are a valid way of detecting planetary mass bodies. Eclipse timing variations are particularly useful for detecting planetary mass bodies in orbital configurations that would go undetected with other methods such as searching for transits which require specific orbital characteristics (such as a compatible inclination) to be viewed from Earth. The J-H magnitude difference of KIC7821010 from 2MASS is 0.195 and approximately matches the J-H magnitude difference of a 1.25$M_\odot$ F6V star which is consistent with the mass estimated for the primary star of the system. A planetary mass third body would contribute essentially nothing to the colours of the system and therefore allows for more accurate estimates of the masses of the primary and secondary stars. 

\subsection{KIC11519226} \label{kic11519226}

KIC11519226 comprises an outer companion with a mass of 1.25$M_\odot$ \citep{borkovits2016}, and eclipsing binary primary and secondary star mass of 0.98$M_\odot$ and 0.46$M_\odot$, respectively. Like KIC5653126 in section \ref{kic5653126}, a third body with such a large mass relative to the binary stars would dominate the light from the system.

\begin{figure}
    \includegraphics[width=\columnwidth]{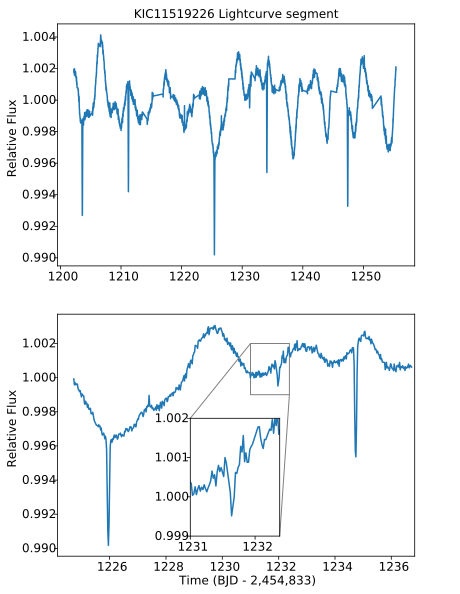}
    \caption{Top Panel: A segment of the light curve of KIC11519226 showing a number of primary and secondary eclipses, the variability in the light curve and a possible extra eclipsing event. Bottom Panel: A possible extra eclipsing event in the light curve of KIC11519226. The primary eclipse can be seen on the left, the secondary eclipse on the right and the possible extra eclipsing event is shown in the rectangle. Given the long period of the outer companion, no additional eclipses would be seen and secondary eclipses are likely lost in the variability of the light curve itself.}
    \label{fig:extra11519226event}
\end{figure}

The inclination of 89\textdegree\ for an additional body around KIC11519226 indicates the possibility of additional eclipse events taking place within the light curve, however there is a lot of variability within the light curve of KIC11519226 which could hide such events. The long period nature of the additional bodies would also limit the number of eclipses that could be observed. Period04 \citep{lenz2005} was used to attempt to clean the periodicity from the light curve of KIC11519226 in an attempt to locate additional eclipsing events without success. Despite this, there is a possible additional eclipse event located within the light curve as seen in Fig. ~\ref{fig:extra11519226event}, however more observations would be required to confirm if this is an additional eclipse or some other kind of variability. 

The 2MASS J-H magnitude difference of KIC11519226 is 0.321 which is approximately equivalent to a G6V star and closely matches the estimate for the primary star. This J-H colour, coupled with the possibility of an additional shallow eclipsing event despite the 89\textdegree\ inclination, suggests that similar to KIC5653126 the reported third body may contribute nothing to the colours of the system. An outer companion with a larger mass than the primary and secondary star that doesn't contribute to the colour of the system suggests the outer companion may be an additional binary, comprised of two smaller stars, or a white dwarf.

\begin{figure}
    \includegraphics[width=\columnwidth]{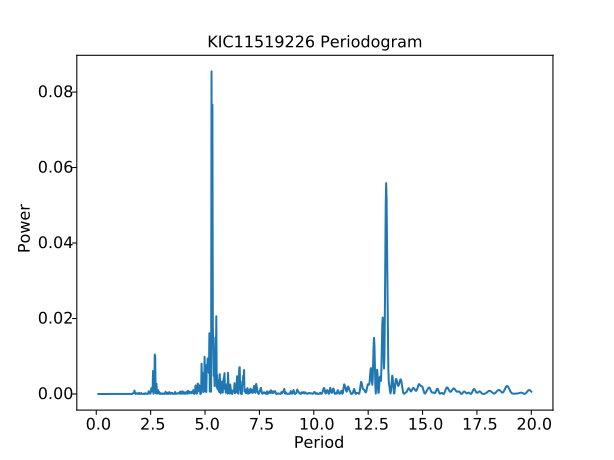}
    \caption{A periodogram of the variability in the out of eclipse light curve of KIC11519226. Two large peaks can be seen, one at 5.3023 days and one at 13.3276 days and one smaller peak can be seen at 2.7084 days.}
    \label{fig:11519226_periodogram}
\end{figure}

A periodogram of the variability was produced using the Lomb-Scargle approach in {\sc gatspy} \citep{vanderplas2015} and is shown in Fig. ~\ref{fig:11519226_periodogram}. Two large peaks can be seen, the first at 5.3023 days and the second at 13.3276 days while a smaller peak can be seen at 2.7084 days. The variability periods of 2.7084 days, 5.3023 days and 13.3276 days are in an approximately 1:2:5 ratio.

$\delta$ scuti variable stars exhibit pulsations in the orders of hours \citep{rodriguez2001} while $\gamma$ Doradus variable stars are typically early F- to late A-type stars \citep{vanreeth2016} as opposed to the G6 primary star estimated in this system. The vast majority of $\gamma$ Doradus candidates listed in \cite{handler1999} have variability periods of less than 2 days. One system, HD 109838, stands out as an exception to the typical periods of a $\gamma$ Doradus star with possible periods of 14 days and 2.9 days however the periods are listed as uncertain. The variability periods for HD 109838 are comparable to the variability periods seen in KIC11519226.

\section{Summary and Conclusions}

In this study we used custom software {\sc bet} based on {\sc transit analysis package} or {\sc tap} to perform an eclipse timing study on Kepler eclipsing binary stars. During the eclipse timing study we found systems that had O-C diagrams that displayed "flip-flop" or out of phase variations between the primary and secondary eclipse O-C curves and rapid period change variations. {\sc rebound} was used to simulate these systems. The systems in Table ~\ref{tab:systems_shortperiod} and Table ~\ref{tab:systems_longperiod} were chosen as they all exhibited a unique "flip-flop" effect within their O-C diagrams. Outer companions with the characteristics described all account for the features seen in the O-C diagrams such as the out of phase eclipse time variations and the "flip-flop" effect. With the systems simulated in {\sc rebound} we then integrated the systems as described for $10^6$ years. We found that all systems were dynamically stable for at least $10^6$ years and, therefore, bodies in these orbital configurations are likely to be stable and observable. We also integrated these systems with random values for the mean longitude, argument of pericentre and longitude of pericentre of the third body and the binary star for $10^4$ years and found that the systems were stable for a wide range of orbital configurations. The evidence suggests the outer companions for the systems listed in Table ~\ref{tab:systems_shortperiod} are an additional pair of stars in a binary configuration (KIC5255552, KIC5653126 and KIC11519226), a single M Dwarf star (KIC5731312 and KIC8023317) and a planet (KIC7821010).

We also suspect that a larger number of systems that have been observed would also show similar "flip-flop" characteristics if observed over longer or much longer time spans. However, due to the limits of the Kepler viewing window and large orbital periods estimated for the third bodies/outer companions the "flip-flop" effect continues to go undetected. As more and more systems are found with multiple bodies, the dynamical stability of the system as a whole is an important consideration when determining the likelihood of their existence. Of particular note is KIC7821010 which has a third body mass of $\sim$2.6 Jupiter masses. At $\sim$2.6 Jupiter masses it is well within planetary mass range and shows that even a relatively small mass can have large effects on the motion of its parent stars.

Other stand-out systems from this study include KIC5255552, where there are additional eclipses in the light curve \citep{zhang2018} that may indicate the presence of a fourth star bound in a binary with the third star. A fifth body in the KIC5255552 system is a possibility and further observations of the system are crucial in determining the true nature of this system. While a triple star explanation cannot be ruled out for the systems KIC5653126 and KIC11519226, the photometric and dynamical analysis performed for this study suggests these systems are detached eclipsing binary stars with binary star companions.

Some of the systems presented, for example KIC11558882, cannot be reliably studied with ground based observations. The orbital period of the binary stars can be so great that observing eclipses to get meaningful data was only made possible with Kepler. Without space based observations these systems, and their O-C variations, may have continued to go undetected.

TESS \citep{ricker2014} is an all-sky survey of bright local stars with the ability of detecting planets with orbital periods of a few hours to a year or more. The launch of TESS provides more opportunities to locate comparable systems that are more local to the solar system and capable of follow up studies. With the launch of TESS and future projects we expect the number of systems that have similar characteristics to increase significantly.

\section*{Acknowledgements}

This research has been supported by an Australian Government Research Training Program Scholarship. We would like to thank the referee and editor for their helpful feedback.

\section*{Data availability}
The data underlying this article will be shared on reasonable request to the corresponding author.



\bibliographystyle{mnras}
\bibliography{getley} 


\bsp	
\label{lastpage}
\end{document}